\documentclass[preprint,superscriptaddress,showpacs,amsmath,amssymb,aps,floats,a4paper]{revtex4}
\usepackage{bm}
\usepackage{graphicx}
\newcommand{\beq}{\begin{equation}}
\newcommand{\eeq}{\end{equation}}

\newcommand{\bfb}{\mbox{\boldmath $b$}}
\newcommand{\bff}{\mbox{\boldmath $f$}}

\newcommand{\bfh}{\mbox{\boldmath $h$}}
\newcommand{\bfk}{\mbox{\boldmath $k$}}
\newcommand{\bfu}{\mbox{\boldmath $u$}}
\newcommand{\bfv}{\mbox{\boldmath $v$}}
\newcommand{\bfx}{\mbox{\boldmath $x$}}
\newcommand{\bfr}{\mbox{\boldmath $r$}}
\newcommand{\bfB}{\mbox{\boldmath $B$}}
\newcommand{\bfJ}{\mbox{\boldmath $J$}}
\newcommand{\bfH}{\mbox{\boldmath $H$}}
\newcommand{\bfK}{\mbox{\boldmath $K$}}

\newcommand{\bfX}{\mbox{\boldmath $X$}}

\newcommand{\bfxi}{\mbox{\boldmath $\xi$}}
\newcommand{\ex}{\mbox{{\boldmath $e$}}_{1}}
\newcommand{\ey}{\mbox{{\boldmath $e$}}_{2}}
\newcommand{\ez}{\mbox{{\boldmath $e$}}_{3}}
\newcommand{\bfemf}{\mbox{\boldmath ${\cal E}$}}
\newcommand{\bnabla}{\mbox{\boldmath $\nabla$}}
\newcommand{\cross}{\mbox{\boldmath $\times$}}
\newcommand{\cendot}{\mbox{\boldmath $\cdot\,$}}
\newcommand{\rem}{{\rm Rm}}
\newcommand{\re}{{\rm Re}}

\begin{document}

\title{Transport coefficients for the shear dynamo problem at small Reynolds numbers}
\author{Nishant K. Singh}\email{nishant@rri.res.in}
\affiliation{
Raman Research Institute, Sadashivanagar, Bangalore 560 080, India}
\affiliation{
Joint Astronomy Programme, Indian Institute of Science, Bangalore 560 012, India}
\author{S. Sridhar}\email{ssridhar@rri.res.in}
\affiliation{
Raman Research Institute, Sadashivanagar, Bangalore 560 080, India}
\date{\today}

\begin{abstract}
We build on the formulation developed in Sridhar \& Singh~({\emph{JFM}, {\bf 664}, 265, 2010}), and present a theory of the \emph{shear dynamo problem} for small magnetic and fluid Reynolds numbers, but for arbitrary values of the shear parameter.  Specializing to the case of a mean magnetic field that is slowly varying in time, explicit expressions for the transport coefficients, $\alpha_{il}$ and $\eta_{iml}$, are derived. We prove that, when the velocity field is non helical, the transport coefficient  $\alpha_{il}$ vanishes. We then consider forced, stochastic dynamics for the incompressible velocity field at low Reynolds number. An exact, explicit solution for the velocity field is derived, and the velocity spectrum tensor is calculated in terms of the Galilean--invariant forcing statistics. We consider forcing statistics that is non helical, isotropic and delta--correlated--in--time, and specialize to the case when the mean--field is a function only of the spatial coordinate $X_3$ and time $\tau\,$; this reduction is necessary for comparison with the numerical experiments of Brandenburg, R{\"a}dler, Rheinhardt \& K\"apyl\"a (\emph{ApJ}, {\bf 676}, 740, 2008). Explicit expressions are derived for all four components of the magnetic diffusivity tensor, $\eta_{ij}(\tau)\,$. These are used to prove that the shear--current effect cannot be responsible for dynamo action at small $\re$ and $\rem$, but for all values of the shear parameter. 
\end{abstract}


\maketitle

\section{Introduction}

Astrophysical systems like planets, galaxies and clusters of galaxies possess magnetic fields which exhibit definite spatial ordering, in addition to a random component. The ordered (or ``large--scale'') components are thought to originate from turbulent dynamo action in the electrically conducting fluids 
in these objects. The standard model of such a process involves amplification of seed magnetic fields due to turbulent flows which lack mirror--symmetry (equivalently, which possess helicity) \cite{M78,KR80,BS05}. The turbulent flows generally possess large--scale shear, which is expected to have significant effects on transport properties \cite{LK09}; however, it is not clear whether the turbulent flows are always helical. Recent work has explored the possiblity that non--helical turbulence in conjunction with background shear may give rise to large--scale dynamo action \cite{BRRK08,Yousef08a,Yousef08b,KKB,HP09,RK03}. The evidence for this comes mainly from direct numerical simulations \cite{BRRK08,Yousef08a, Yousef08b}, but it by no means clear what physics drives such a dynamo.  One possibility that has received some attention is the shear--current effect \cite{RK03}, where an extra component of the mean electromotive force (EMF) is thought to result in the generation of the cross--shear component of the mean magnetic field
from the component parallel to the shear flow. However, there is no agreement yet whether the sign of such a coupling is favorable to the operation of a dynamo; some analytic calculations \cite{RS06,RK06} and numerical experiments \cite{BRRK08} find that the sign of the shear--current term is unfavorable for dynamo action. 

A quasilinear kinematic theory of dynamo action in a linear shear flow of an incompressible fluid which has random velocity fluctuations was presented in \cite{SS09}, who used the ``second order correlation approximation'' (SOCA) in the limit of zero resistivity. Unlike earlier analytic work which treated shear as a small perturbation, this theory did not place any restriction on the strength of the shear. They arrived at an integro--differential equation for the evolution of the mean magnetic field and argued that the shear-current assisted dynamo is essentially absent. The theory was extended to take account of non zero resistivity in \cite{SS10}; this is again nonperturbative in the shear strength, uses SOCA, and is rigorously valid in the limit of small magnetic Reynolds number ($\rem$) but with no restriction on the fluid Reynolds number ($\re$). The kinematic approach to the \emph{shear dynamo problem} taken in \cite{SS09, SS10} uses in an essential manner the shearing coordinate transformation and the Galilean invariance (which is a fundamental symmetry of the problem) of the velocity fluctuations. The present work extends \cite{SS10} by giving definite form to the statistics of the velocity field; specifically, the velocity field is assumed to obey the forced Navier--Stokes equation, in the absence of Lorentz forces. 

In section~II we begin with a brief review of the salient results of \cite{SS10}. The expression for the 
Galilean--invariant mean EMF is then worked out for the case of a mean magnetic field that is slowly varying in time. 
Thus the mean--field induction equation, which is an integro--differential equaton in the formulation of \cite{SS10}
now simplifies to a partial differential equation. This reduction is an essential first step to the later comparision with 
the numerical experiments of \cite{BRRK08}. Explicit expressions for the transport coefficients, $\alpha_{il}$ and 
$\eta_{iml}$, are derived in terms of the two--point velocity correlators. We then recall some results from \cite{SS10}, 
which express the velocity correlators in terms of the velocity spectrum tensor. This tensorial quantity is real when the velocity 
field is non helical; we are able to prove that, in this case,  the transport coefficient  $\alpha_{il}$ vanishes. Section~III
develops the dynamics of the velocity field at low $\re$, using the Navier--Stokes equation with stochastic external forcing. 
An explicit solution for the velocity field is presented and the velocity spectrum tensor is calculated in terms of the 
Galilean--invariant forcing statistics. For non helical forcing, the velocity field is also non helical and the transport coefficient  
$\alpha_{il}$ vanishes, as noted above. We then specialize to the case when the forcing is not only non helical, but isotropic 
and delta--correlated--in--time as well. In section~IV we specialize to the case when the mean--field is a function only of the spatial coordinate $X_3$ 
and time $\tau\,$; this reduction is necessary for comparision with the numerical experiments of \cite{BRRK08}. Explicit 
expressions are derived for all four components of the magnetic diffusivity tensor, $\eta_{ij}(\tau)\,$, in terms of the 
velocity power spectrum; the 
late--time saturation values, $\eta^\infty_{ij}\,$, have direct bearing on the growth (or otherwise) of the mean magnetic field. 
Comparisons with earlier work---in particular  \cite{BRRK08}---are made, and the implications for the shear--current effect 
are discussed. We then conclude in section~V.

\section{Mean--field electrodynamics in a linear shear flow}

\subsection{Mean--field induction equation for small $\rem$}

We begin with a brief review of the main results of \cite{SS10}.
Let $(\ex,\ey,\ez)$ be the unit basis vectors of a Cartesian coordinate system in the laboratory frame. Using notation,  $\bfX = (X_1,X_2,X_3)$  
for the position vector and $\tau$ for time, we write the  fluid velocity as $(SX_1\ey + \bfv)$, where $S$ is the rate of shear parameter 
and $\bfv(\bfX, \tau)$ is an incompressible and randomly fluctuating velocity field with zero mean. The mean magnetic field, $\bfB(\bfX, \tau)$, 
obeys the following (mean--field induction) equation:

\beq
\left(\frac{\partial}{\partial\tau} \;+\; SX_1\frac{\partial}{\partial X_2}\right)\bfB \;-\; SB_1\ey \;=\; 
\bnabla\cross\bfemf \;+\; \eta\bnabla^2\bfB
\label{meanindeqn}
\eeq

\noindent
where $\eta$ is the microscopic resistivity, and  $\bfemf$ is the mean electromotive force (EMF), $\bfemf = \left<\bfv\cross\bfb\right>$, where 
$\bfv$ and $\bfb$ are the fluctuations in the velocity and magnetic fields.
 
To lowest order in $\rem$, the evolution of the magnetic field fluctuations, now denoted by $\bfb^{(0)}$, is governed by,

\beq 
\left(\frac{\partial}{\partial\tau} \;+\; SX_1\frac{\partial}{\partial X_2}\right)\bfb^{(0)} \;-\; Sb_1^{(0)}\ey \;=\; \left(\bfB\cendot\bnabla\right)\bfv \;-\; \left(\bfv\cendot\bnabla\right)\bfB \;+\; \eta\bnabla^2\bfb^{(0)}
\label{flucindlineqn}
\eeq

\noindent
This equation was solved by making a \emph{shearing--coordinate transformation} to new spacetime coordinates and new field variables. The new spacetime variables, $\left(\bfx, t\right)$, are given by,

\beq
x_1 = X_1\,,\qquad x_2 = X_2 - S\tau X_1\,,\qquad x_3 = X_3\,,\qquad t = \tau\,,
\label{sheartr}
\eeq

\noindent
where $\bfx$ may be thought of as the Lagrangian coordinates of a fluid element in the background shear flow. The new field variables are component--wise equal to the old variables: 
\beq
\bfH(\bfx, t) \;=\; \bfB(\bfX, \tau)\,,\qquad
\bfh(\bfx, t) \;=\; \bfb^{(0)}(\bfX, \tau)\,,\qquad
\bfu(\bfx, t) \;=\; \bfv(\bfX, \tau)
\label{newvar}
\eeq

\noindent
In the new variables, equation~(\ref{flucindlineqn}) becomes,

\beq
\frac{\partial\bfh}{\partial t} \;-\; Sh_1\ey \;=\; \left(\bfH\cendot\frac{\partial}{\partial\bfx}
- StH_1\frac{\partial}{\partial x_2}\right)\bfu \;-\; \left(\bfu\cendot\frac{\partial}{\partial\bfx} - Stu_1\frac{\partial}{\partial x_2}\right)\bfH \;+\; \eta\bnabla^2\bfh
\label{newvareqn}
\eeq

\noindent
We need the particular solution (i.e. the \emph{forced solution}) which vanishes at $t=0$. This is given in component form as,

\begin{eqnarray}
h_m(\bfx,t) &=& \int_0^t \mathrm{d}t' \int \mathrm{d}^3x' \, G_{\eta}(\bfx-\bfx',t,t')\,
\left[u'_{ml} + S(t-t')\delta_{m2} u'_{1l}\right] \times \nonumber \\[2ex]
&& \qquad\qquad\qquad\qquad\qquad \times \left[H'_l - St'\delta_{l2}H'_1\right]\nonumber\\[3ex]
&-& \int_0^t \mathrm{d}t'\int \mathrm{d}^3x'\,G_{\eta}(\bfx-\bfx',t,t')\,
\left[H'_{ml} + S(t-t')\delta_{m2}H'_{1l}\right] \times \nonumber \\[2ex]
&& \qquad\qquad\qquad\qquad\qquad \times \left[u'_l - St'\delta_{l2}u'_1\right]
\label{hsoln}
\end{eqnarray}

\noindent
The primes in $H'_l$ and $u'_l$ mean that these functions are evaluated at $\left(\bfx', t'\right)$. The quantities $H'_{ml}$ and $u'_{ml}$ are shorthand for $\left(\partial H'_m/\partial x'_l\right)$ and $\left(\partial u'_m/\partial x'_l\right)$, respectively. Here  $G_{\eta}(\bfr,t,t')$  is the \emph{resistive Green's function} for the linear shear flow \cite{KR71, SS10}, which takes the form of a sheared heat kernel. The one property we will use is that $G_{\eta}(\bfr,t,t')$ is an even function of $\bfr$. 
Otherwise, its spatial Fourier transform, defined by

\begin{eqnarray}
\widetilde{G}_{\eta}(\bfk,t,t') &=& \int \mathrm{d}^3r \,  G_{\eta}(\bfr,t,t')\,\exp{\left[-\mathrm{i}\,\bfk\cendot\bfr\right]}\nonumber\\[2ex]
&=& \exp{\left[-\eta\left(k^2(t-t') - S\,k_1\,k_2(t^2-t^{\prime 2}) + \frac{S^2}{3}\,k_2^2 (t^3-t^{\prime 3}) \right)\right]}
\label{frgreta}
\end{eqnarray}

\noindent
is more useful for our purposes.

The mean EMF is given by $\bfemf = \left<\bfv\cross\bfb^{(0)}\right> = \left<\bfu\cross\bfh\right>$, where equation~(\ref{hsoln}) for $\bfh$ should be substituted. The averaging, $\left<\;\;\right>$, acts only on the velocity variables but not the mean field; i.e. $\left<\bfu\bfu\bfH\right> = \left<\bfu\bfu\right>\bfH$ etc. The $\bfu\bfu$ velocity correlators can be rewritten in terms of the $\bfv\bfv$ velocity correlators; this is a useful step because the latter are referred to the laboratory frame. The velocity correlators have a very important property called \emph{Galilean invariance}, which is shared by \emph{comoving observers}, who translate uniformly with the background shear flow. If a comoving observer is at position $\bfxi = (\xi_1, \xi_2, \xi_3)$ at the initial time, then at a later time $t$, her 
location is given by,  
 
\beq  
\bfX_c \left(\bfxi,t \right) \;=\; \left(\xi_1, \xi_2 + St\,\xi_1, \xi_3\right)
\label{comove}
\eeq

\noindent
Velocity fluctuations are defined to be Galilean--invariant, if and only if the statistical properties of the fluctuations, as seen by any 
comoving observer,  are identical to the statistical properties seen in the laboratory frame; it follows that all comoving observers see the
 same statistics. There are two basic Galilean--invariant two--point velocity correlation functions, $Q_{jml}$ and $R_{jm}$, which are defined as:

\begin{eqnarray}
Q_{jml}(\bfr,t,t') &\;=\;& \left< v_j \left(\bfX_c \left(\frac{\bfr}{2},t \right),t \right) \frac{\partial v_m}{\partial X_l} 
\left(\bfX_c \left(-\frac{\bfr}{2},t' \right),t' \right) \right>\nonumber\\[3ex]
R_{jm}(\bfr,t,t') &\;=\;& \left< v_j \left(\bfX_c \left(\frac{\bfr}{2},t \right),t \right) v_m 
\left(\bfX_c \left(-\frac{\bfr}{2},t' \right),t' \right) \right>
\label{qrdef}
\end{eqnarray}

\noindent
Then the mean EMF is a \emph{functional} of the mean magnetic field, $H_l$, and its first spatial derivative, $H_{lm} = (\partial H_l/\partial x_m)$:

\begin{eqnarray}
{\cal E}_i(\bfx, t)
\;=\; && \epsilon_{ijm}\,\int_0^t \mathrm{d}t' \int \mathrm{d}^3r \; G_{\eta}(\bfr,t,t')\,C_{jml}(\bfr,t,t')H_l(\bfx-\bfr,t') \;-\; \nonumber\\[3ex]
\;&-&\; \int_0^t \mathrm{d}t' \int \mathrm{d}^3r \, G_{\eta}(\bfr,t,t')\left[\epsilon_{ijl} + S(t-t')\delta_{l1}\epsilon_{ij2}\right] 
D_{jm}(\bfr,t, t')H_{lm}(\bfx-\bfr,t')\,.\nonumber \\
&&\label{emfcd}
\end{eqnarray}

\noindent 
where $C_{jml}$ and $D_{jm}$ are two--point velocity correlators, which are derived from the more basic two--point velocity correlators, $Q_{jml}$ and $R_{jm}$, of equations~(\ref{qrdef}):

\begin{eqnarray}
C_{jml}(\bfr,t,t') &\;=\;& Q_{jml}(\bfr,t,t') \;+\; S(t-t')\delta_{m2}\,Q_{j1l}(\bfr,t,t')\,,\nonumber\\[2ex]
D_{jm}(\bfr,t,t') &\;=\;& R_{jm}(\bfr,t,t') \;-\; St'\delta_{m2}\,R_{j1}(\bfr,t,t')\,.
\label{cddef}
\end{eqnarray}

\noindent
Then the time evolution of the mean magnetic field is given in the new variables by,

\begin{eqnarray}
\frac{\partial\bfH}{\partial t} \;-\; SH_1\ey &\;=\;& \bnabla\cross\bfemf \;+\; \eta\bnabla^2\bfH\,,\nonumber\\[2ex]
\left(\bnabla\right)_p \;\equiv\; \frac{\partial}{\partial X_p} &\;=\;& 
\frac{\partial}{\partial x_p} \;-\; St\,\delta_{p1}\frac{\partial}{\partial x_2}\,,
\label{meanindeqnH}
\end{eqnarray}

\noindent
Equations~(\ref{meanindeqnH}) and (\ref{emfcd}) form a closed system of integro--differential equations, determining the time evolution of the mean magnetic field, $\bfH(\bfx, t)$. 

\subsection{The mean EMF for a slowly varying magnetic field}

The mean EMF given in equation~(\ref{emfcd}) is a \emph{functional} of $H_l$ and $H_{lm}$. 
When the mean--field is slowly varying compared to velocity correlation times, we expect to be able to approximate 
$\bfemf$ as a \emph{function} of $H_l$ and $H_{lm}$. In this case, the mean--field induction equation would reduce to 
a set of coupled partial differential equations, instead of the more formidable set of coupled integro--differential 
equations given by (\ref{meanindeqnH}) and (\ref{emfcd}). Sheared coordinates are essential for the calculations, but 
physical interpretation is simplest in the laboratory frame; hence we derive an expression for the mean EMF in terms of
$\bfB(\bfX, \tau)$.

The first step involves a  Taylor expansion of the quantitites, $H_l$ and $H_{lm}$, occuring in equation~(\ref{emfcd}) for the mean EMF. Neglecting spacetime derivatives higher than the first order ones, we have  

\begin{eqnarray}
H_l(\bfx-\bfr,t') &=& H_l(\bfx,t) \;-\; r_p H_{lp}(\bfx,t) \;-\; (t-t')\,\frac{\partial H_l (\bfx,t)}{\partial t} \;+\; \ldots\nonumber\\[2ex]
H_{lm}(\bfx-\bfr,t') &=& H_{lm}(\bfx,t) \;-\; (t-t')\,\frac{\partial H_{lm} (\bfx,t)}{\partial t} \;+\;  \ldots
\label{texph}
\end{eqnarray}

\noindent
We now use the mean--field induction equation~(\ref{meanindeqnH}), to express $\left(\partial\bfH/\partial t\right)$ in terms of spatial derivatives.
Let $L$ be the spatial scale over which the mean--field varies. When the mean--field varies slowly, $L$ is large and the contributions from both the resistive term and the mean EMF are small, as is shown below. Let $\ell$ and $v_{\rm rms}$ be the spatial scale and root--mean--squared amplitude of the velocity fluctuations. The resistive term makes a contribution of order $(\ell/L)^2\rem^{-1}$, which we now assume is much less than unity. Using equation~(\ref{emfcd}), we can verify that $\bnabla\cross\bfemf$ contributes terms of five different orders; $(\ell/L)$, $(\ell/L)(S\ell/v_{\rm rms})$, $(\ell/L)^2$, $(\ell/L)^2(S\ell/v_{\rm rms})$ and $(\ell/L)^2(S\ell/v_{\rm rms})^2$. These are all small when $(\ell/L)$ and $(\ell/L)(S\ell/v_{\rm rms})$ are both much smaller than unity. That we must have $(\ell/L)\ll 1$ is natural from the familiar case of zero shear. The presence of shear introduces an additional requirement that $(\ell/L)(S\ell/v_{\rm rms})\ll 1$. We now define the small parameter, $\mu\ll 1$, to be equal to the largest of the three small quantities, $(\ell/L)^2\rem^{-1}\ll 1$,  $(\ell/L)\ll 1$ and $(\ell/L)(S\ell/v_{\rm rms})\ll 1$. Then,

\beq
\frac{\partial H_l}{\partial t} \;=\; S\delta_{l2}H_1 \;+\; O(\mu)
\label{indpert}
\eeq

\noindent
and equations~(\ref{texph}) give,  

\begin{eqnarray}
H_l(\bfx-\bfr,t') &=& H_l(\bfx,t) \;-\; r_p H_{lp}(\bfx,t) \;-\; S(t-t')\delta_{l2}H_1 \;+\; O(\mu)\nonumber\\[2ex]
H_{lm}(\bfx-\bfr,t') &=& H_{lm}(\bfx,t) \;-\; S(t-t')\delta_{l2}H_{1m} \;+\; O(\mu)
\label{texpheps}
\end{eqnarray}

\noindent
We substitute equation~(\ref{texpheps}) in (\ref{emfcd}) to get, 

\begin{eqnarray}
{\cal E}_i(\bfx, t) &=& \epsilon_{ijm}\,H_l(\bfx, t)\int_0^t \mathrm{d}t' \int \mathrm{d}^3r \, G_{\eta}(\bfr,t,t')\;
\left[C_{jml}(\bfr,t,t')\,-\,S(t-t')\,\delta_{l1}\,C_{jm2}(\bfr,t,t')\right]\nonumber\\[3ex]
&&\quad -\,\epsilon_{ijm}\,H_{lp}(\bfx, t)\int_0^t \mathrm{d}t' \int \mathrm{d}^3r \; r_p\,G_{\eta}(\bfr,t,t')\;
C_{jml}(\bfr,t,t')\nonumber\\[3ex]
&&\quad-\,\epsilon_{ijl}\,H_{lm}(\bfx, t)\int_0^t \mathrm{d}t' \int \mathrm{d}^3r \,G_{\eta}(\bfr,t,t')\;D_{jm}(\bfr,t,t')
\;+\; O(\mu^2)
\label{emftemp}
\end{eqnarray}

\noindent
The final step is to rewrite the above expression in terms of the original magnetic field variable, using,

\begin{eqnarray}
H_l(\bfx, t) &=& B_l(\bfX, \tau)\nonumber \\[2ex]
H_{lm}(\bfx,t) &\equiv & \frac{\partial H_l(\bfx,t)}{\partial x_m} \;=\;
\left(\frac{\partial}{\partial X_m}\,+\,S\tau\delta_{m1}\frac{\partial}{\partial X_2} \right)\,B_l(\bfX, \tau)
\label{hbtr}
\end{eqnarray}

\noindent
Therefore, for a slowly varying magnetic field, the mean EMF is given by, 

\beq
{\cal E}_i \;=\; \alpha_{il}(\tau) B_l(\bfX,\tau) \;-\; \eta_{iml}(\tau)\,\frac{\partial B_l(\bfX,\tau)}{\partial X_m}
\label{emfslow}
\eeq

\noindent
where the \emph{transport coefficients} are given by,

\begin{eqnarray}
\alpha_{il}(\tau)\;&=&\;\epsilon_{ijm}\, \int_0^{\tau} \mathrm{d}\tau' \int \mathrm{d}^3r \, G_{\eta}(\bfr,\tau,\tau')\,
\left[C_{jml}(\bfr,\tau,\tau')\,-\,S(\tau-\tau')\,\delta_{l1}\,C_{jm2}(\bfr,\tau,\tau')\right]\nonumber \\ [3ex]
\eta_{iml}(\tau)\;&=&\; \epsilon_{ijp}\,\int_0^{\tau} \mathrm{d}\tau' \int \mathrm{d}^3r \;[r_m+S\tau\delta_{m2}r_1]\,
G_{\eta}(\bfr,\tau,\tau')\,C_{jpl}(\bfr,\tau,\tau') \;+\; \nonumber \\ [2ex]
&&+\; \epsilon_{ijl}\,\int_0^{\tau} \mathrm{d}\tau' \int \mathrm{d}^3r \, G_{\eta}(\bfr,\tau,\tau')\;
[D_{jm}(\bfr,\tau, \tau')+S\tau\delta_{m2}D_{j1}(\bfr,\tau,\tau')] 
\label{trcoef}
\end{eqnarray}

\noindent
Then the mean--field induction equation~(\ref{meanindeqn}), together with equations~(\ref{emfslow}) and (\ref{trcoef}), is a closed partial differential equation (which is first order in temporal and second order in spatial derivatives). 

\subsection{Velocity correlators expressed in terms of the velocity spectrum tensor}
The Galilean invariance of the two--point velocity correlators can be stated most compactly in Fourier--space. 
Let $\tilde{\bfv}(\bfK, \tau)$ be the spatial Fourier transform of $\bfv(\bfX, \tau)$, defined by

\beq
\tilde{\bfv}(\bfK, \tau) \;=\; \int \, \mathrm{d}^3X \,\bfv(\bfX, \tau)\,\exp{\left[-\mathrm{i} \bfK\cendot\bfX\right]} \; ; \qquad \qquad
\left[\bfK\cendot\tilde{\bfv}(\bfK,\tau) \right] \,=\, 0 
\label{ftv1}
\eeq

\noindent
New Fourier variables are defined by the Fourier--space \emph{shearing transformation}, 

\beq
k_1\,=\,K_1\,+\,S\tau K_2, \qquad k_2\,=\,K_2, \qquad k_3\,=\,K_3, \qquad t\,=\,\tau
\label{shtr}
\eeq

\noindent
It is proved in \cite{SS10} that a Galilean--invariant Fourier--space two--point velocity correlator must be of the form

\beq
\left<\tilde{v}_j(\bfK, \tau)\,\tilde{v}_m^*(\bfK', \tau')\right> \;=\; 
(2\pi)^6\,\delta(\bfk -\bfk')\,\Pi_{jm}(\bfk, t, t')
\label{spec}
\eeq

\noindent
where $\Pi_{jm}$ is the {\it velocity spectrum tensor}, which must possess the following properties:
 
\begin{eqnarray}
\Pi_{ij}(\bfk, t, t') &\;=\;& \Pi_{ij}^*(-\bfk, t, t') \;=\; \Pi_{ji}(-\bfk, t', t)
\nonumber\\[2ex]
K_i\,\Pi_{ij}(\bfk, t, t') &\;=\;& \left(k_i - St\,\delta_{i1}k_2\right)\Pi_{ij}(\bfk, t, t') \;=\; 0
\nonumber\\[2ex]
K'_j\,\Pi_{ij}(\bfk, t, t') &\;=\;& \left(k_j - St'\,\delta_{j1}k_2\right)\Pi_{ij}(\bfk, t, t') \;=\; 0
\label{specprop}
\end{eqnarray}

Now, the various two--point velocity correlators can be written as:

\begin{eqnarray}
R_{jm}(\bfr,t,t') &=& \int\,\mathrm{d}^3k\,\Pi_{jm}(\bfk,t,t')\,\exp{\left[\mathrm{i}\,\bfk\cendot\bfr\right]}\nonumber\\[3ex]
Q_{jml}(\bfr,t,t') &=& -\mathrm{i}\,\int\,\mathrm{d}^3k\,\left[k_l - St'\delta_{l1}k_2\right]\,\Pi_{jm}(\bfk,t,t')\,
\exp{\left[\mathrm{i}\,\bfk\cendot\bfr\right]}\nonumber\\[3ex]
D_{jm}(\bfr,t,t') &=& \int\,\mathrm{d}^3k\,\left[\Pi_{jm}(\bfk,t,t')  - St'\delta_{m2}\,\Pi_{j1}(\bfk,t,t')\right]\exp{\left[\mathrm{i}\,
\bfk\cendot\bfr\right]}\nonumber\\[3ex]
C_{jml}(\bfr,t,t') &=& -\mathrm{i}\,\int \mathrm{d}^3k\left[k_l - St'\delta_{l1}k_2\right]\left[\Pi_{jm}(\bfk,t,t') +  S(t-t')
\delta_{m2}\Pi_{j1}(\bfk,t,t')\right]\exp{\left[\mathrm{i}\,\bfk\cendot\bfr\right]}\nonumber\\
&&\label{rqdc}
\end{eqnarray}

\noindent
Using the above expressions for $D_{jm}$ and $C_{jml}$ in equations~(\ref{trcoef}), the transport coefficients 
$\alpha_{il}(\tau)$ and $\eta_{iml}(\tau)$ can also be written in terms of the velocity spectrum tensor. 

The \emph{correlation helicity} may be defined as,

\beq
H_{\rm cor}(t,t') \;=\; \epsilon_{jlm}\left<v_j({\bf 0},t)\,v_{ml}({\bf 0},t')\right>
\;=\; \mathrm{i}\,\int \mathrm{d}^3k\,\left[k_l - St'\delta_{l1}k_2\right]\epsilon_{ljm}\Pi_{jm}(\bfk, t, t')
\label{hcordef} 
\eeq

\noindent
From the first of equations~(\ref{specprop}), it is clear that the real part of $\Pi_{jm}(\bfk, t, t')$ is an even function of $\bfk$, whereas the imaginary part is an odd function of $\bfk$. Hence only the imaginary part
of $\Pi_{jm}(\bfk, t, t')$ contributes to the correlation helicity. We shall see that the forced velocity fields we deal with later in this article possess a real velocity spectrum, and their correlation helicity vanishes. In this case, 

\begin{eqnarray}
Q_{jml}(\bfr,t,t') &=& \int\,\mathrm{d}^3k\,\left[k_l - St'\delta_{l1}k_2\right]\,\Pi_{jm}(\bfk,t,t')\,
\sin{\left[\bfk\cendot\bfr\right]}\nonumber\\[3ex]
C_{jml}(\bfr,t,t') &=& \int \mathrm{d}^3k\left[k_l - St'\delta_{l1}k_2\right]\left[\Pi_{jm}(\bfk,t,t') +  S(t-t')
\delta_{m2}\Pi_{j1}(\bfk,t,t')\right]\sin{\left[\bfk\cendot\bfr\right]}\nonumber\\
&&\label{qcodd}
\end{eqnarray}

\noindent
are both \emph{odd} functions of $\bfr$. Since the resistive Green's function, $G_{\eta}(\bfr,t,t')$, is an \emph{even} function of $\bfr$, 
equation~(\ref{trcoef}) implies that the \emph{transport coefficient $\alpha_{il}(\tau)$ vanishes}.

\section{Forced Stochastic velocity dynamics}

\subsection{Forced velocity dynamics for small $\re$}

We consider the simplest of dynamics for the velocity field by ignoring Lorentz forces, and assuming that the  fluid is stirred 
randomly by some external means. If the velocity fluctuations have root--mean--squared (rms) amplitude $v_{{\rm rms}}$ 
on some typical spatial scale $\ell$,  the fluid  Reynolds number may be defined as $\re = (v_{{\rm rms}}\ell/\nu)$, where $\nu$ is the 
kinematic viscosity; note that $\re$ has been defined with respect to the fluctuation velocity field, not the background shear velocity field. 
In the limit of small Reynolds number ($\re \ll 1$), the nonlinear term in the Navier--Stokes equation may be ignored. Then the dynamics 
of the velocity field, $\bfv(\bfX, \tau)$, is governed by the randomly forced, linearized Navier--Stokes equation, 

\beq
\left(\frac{\partial}{\partial\tau} \;+\; SX_1\frac{\partial}{\partial X_2}\right) \bfv \;+\; Sv_1\ey \,=\, 
- \bnabla p \;+\; \nu \nabla^2 \bfv \;+\; \bff
\label{linfluc}
\eeq

\noindent
$\bff(\bfX, \tau)$ is the random stirring force per unit mass which is assumed to be divergence--free  with zero
mean: $\bnabla\cendot\bff = 0$ and $\left<\bff\right> = {\bf 0}\,$. The pressure variable, $p$, is determined by 
requiring that equation~(\ref{linfluc}) preserves the condition, $\bnabla\cendot\bfv = 0$. Then $p$ satisfies the Poisson equation, 

\beq
\nabla^2 p \;=\; - 2S\frac{\partial v_1}{\partial X_2} 
\label{press}
\eeq

\noindent
It should be noted that the linearity of the equations~(\ref{linfluc}) and (\ref{press}) implies that the velocity fluctuations have zero 
mean, $\left<\bfv\right> = {\bf 0}\,$. It is clear from equation~(\ref{press}) that $p$ is a non local function of the velocity field, so it is 
best to work in Fourier--space. Taking the spatial Fourier transform of equation~(\ref{linfluc}), we can see that the Fourier transform of the 
velocity field, $\tilde{\bfv}(\bfK, \tau)$, obeys, 

\beq
\left( \frac{\partial}{\partial \tau} \;-\; SK_2 \frac{\partial}{\partial K_1}\;+\; \nu K^2 \right) \tilde{v}_i
\;-\; 2S\left( \frac{K_2 K_i}{K^2} \,-\, \frac{\delta_{i2}}{2} \right) \tilde{v}_1\;=\; \tilde{f}_i 
\label{ftlinfluc}
\eeq 
where $\tilde{f}_i(\bfK, \tau)$ is the spatial Fourier transform of $f_i$. It can be verified that the equation~(\ref{ftlinfluc}) preserves the 
incompressibility condition $K_m \tilde{v}_m \;=\; 0$. 

We can get rid of the inhomogeneous term, $\left(K_2 {\partial / \partial K_1}\right)$, in equation~(\ref{ftlinfluc}) by transforming 
from the old variables $(\bfK,\tau)$ to new variables $(\bfk, t)$, through the Fourier--space shearing transformation of equation~(\ref{shtr}). 
First, we need to define new velocity and forcing variables, $a_i(\bfk,t)$ and $g_i(\bfk,t)$, respectively, by

\beq
\tilde{v}_i(\bfK,\tau) \;=\; \widetilde{G}_{\nu}(\bfk,t,0)\,a_i(\bfk,t)
\label{v-a}
\eeq
\beq
\tilde{f}_i(\bfK,\tau) \;=\; \widetilde{G}_{\nu}(\bfk,t,0)\,g_i(\bfk,t)
\label{f-g}
\eeq

\noindent 
where $\widetilde{G}_{\nu}(\bfk,t,0)$ is the Fourier--space {\it  viscous Green's function}, defined by

\beq
\widetilde{G}_{\nu}(\bfk,t,t') \;=\; \exp{\left[ -\nu \int_{t'}^t \, \mathrm{d}s \, K^2(\bfk,s) \right]} \\[1em]
\label{fgrfn1}
\eeq

\noindent Noting the fact that $\bfK(\bfk,s) \,=\, (k_1-Ssk_2, k_2, k_3)$ and $K^2(\bfk,s)\,=\,\lvert \bfK(\bfk,s) \rvert^2$, the 
viscous Green's function can be calculated in explicit form as

\beq
\widetilde{G}_{\nu}(\bfk,t,t') \;=\; \exp{\left[ -\nu \left( k^2(t-t')\,-\,S\,k_1\,k_2\,(t^2-t'^2)\,+\,\frac{S^2}{3}k_2^2
(t^3-t'^3) \right) \right]} \\[1em]
\label{fgrfn2}
\eeq

\noindent 
The Green's function possesses the following properties:

\begin{eqnarray}
&&\widetilde{G}_{\nu}(\bfk,t,t') \;= \; \widetilde{G}_{\nu}(-\bfk,t,t')\nonumber \\[2ex]
&&\widetilde{G}_{\nu}(k_1, k_2, k_3,t,t') \;= \; \widetilde{G}_{\nu}(k_1, k_2, -k_3,t,t')\nonumber \\[2ex]
&&\widetilde{G}_{\nu}(\bfk,t,t') \;=\; \widetilde{G}_{\nu}(\bfk,t,s) \times \widetilde{G}_{\nu}(\bfk,s,t')\,, \qquad\text{for any s}
\label{gfpr1}
\end{eqnarray}

\noindent 
Using the inverse transformation, 

\beq
K_1\,=\,k_1\,-\,Stk_2, \qquad K_2\,=\,k_2, \qquad K_3\,=\,k_3, \qquad \tau\,=\,t
\label{inshtr}
\eeq

\noindent 
and the fact that partial derivatives transform as,

\beq
\frac{\partial}{\partial K_j} \;=\; \frac{\partial}{\partial k_j}\,+\,St\delta_{j2}\frac{\partial}{\partial k_1} \, ;
\qquad  \qquad \frac{\partial}{\partial \tau} \;=\; \frac{\partial}{\partial t}\,+\,Sk_2\frac{\partial}{\partial k_1}
\label{dertr}
\eeq

\noindent 
equation~(\ref{ftlinfluc}) leads to the following equation for the new velocity variables, $a_i(\bfk,t)$:

\beq
\frac{\partial a_i}{\partial t} \;-\;2S\left( \frac{K_2 K_i}{K^2} - \frac{\delta_{i2}}{2} \right)\,a_1 \;=\;g_i \\[1em]
\label{newvelfl}
\eeq

\noindent 
\noindent where $\bfK(\bfk,t) \,=\, (k_1-Stk_2, k_2, k_3)$ and $K^2(\bfk,t)\,=\,\lvert \bfK(\bfk,t) \rvert^2$ as given by equation~(\ref{inshtr}). 
It can be verified that equation~(\ref{newvelfl}) preserves the dot product, $K_ia_i\,=\,0$. We also note that the dependence of the velocities, 
$\tilde{v}_i(\bfK,\tau)$ on the viscosity $\nu$ arises solely through the Fourier--space Green's function. It is helpful to display in explicit form 
all three components of equation~(\ref{newvelfl}):

\begin{eqnarray}
\label{a1}
 \frac{\partial a_1}{\partial t} \;-\;2S\left( \frac{K_1 K_2}{K^2} \right)\,a_1 \;&=&\;g_1 \\[2ex]
\label{a2}
\frac{\partial a_2}{\partial t} \;-\;2S\left( \frac{K_2^2}{K^2} - \frac{1}{2} \right)\,a_1 \;&=&\;g_2 \\[2ex]
\label{a3}
\frac{\partial a_3}{\partial t} \;-\;2S\left( \frac{K_2 K_3}{K^2} \right)\,a_1 \;&=&\;g_3
\end{eqnarray}

\noindent
Then equation~(\ref{a1}) can be solved to get an explicit expression for $a_1(\bfk,t)$.  When this is substituted in 
equations~(\ref{a2}) and (\ref{a3}), they can be integrated directly to obtain expressions for $a_2(\bfk,t)$ and $a_3(\bfk,t)$.
The {\it forced} (or particular) solution, with initial condition $a_i(\bfk,0)\,=\,0$ is

\beq
a_i(\bfk,t) \;\;=\;\; \int_0^t\,\mathrm{d}s\, g_i(\bfk,s) \;+\; \int_0^t\,\mathrm{d}s\, \left[ \Lambda_i(\bfK(\bfk,t))\,-\,\Lambda_i(\bfK(\bfk,s)) \right]\, 
\frac{K^2(\bfk,s)}{K_{\bot}^2}\,g_1(\bfk,s) 
\label{asoln}
\eeq

\noindent 
where $K_{\bot}^2 \,\equiv\, K_2^2 + K_3^2 \,=\ k_2^2 + k_3^2 \,\equiv\, k_{\bot}^2$, and the function, $\Lambda_i$, is 
defined as

\beq
\Lambda_i(\bfK) \;\;=\;\; \delta_{i1} \:-\; \frac{K_1K_i}{K^2} \;+\; \frac{K_3}{K_{\bot}} \left[\frac{K_3}{K_2}\,\delta_{i2} \;-\; \delta_{i3}\right] 
\arctan{\left( \frac{K_1}{K_{\bot}} \right)}
\label{psidef} 
\eeq

\subsection{Velocity spectrum tensor expressed in terms of the forcing}

Our goal is to express the velocity spectrum tensor in terms of the statistical properties of the forcing. If the forcing is Galilean--invariant, then we must have, 

\beq
\left<\tilde{f}_j(\bfK,\tau)\,\tilde{f}_m^*(\bfK',\tau')\right>\;=\;(2\pi)^6\,\delta(\bfk-\bfk')\Phi_{jm}(\bfk,t,t') 
\label{giff}
\eeq

\noindent 
where $\Phi_{jm}$ is the {\it forcing spectrum tensor}. We are now ready to use the dynamical solution of the last subsection. Using equations~(\ref{v-a}) and (\ref{asoln}), Fourier--space, unequal--time, two--point velocity correlator is given by,

\begin{eqnarray}
\left<\tilde{v}_j(\bfK, \tau)\,\tilde{v}_m^*(\bfK', \tau')\right> 
&=& \widetilde{G}_{\nu}(\bfk,t,0)\,\widetilde{G}_{\nu}(\bfk',t',0)\left<\tilde{a}_j(\bfk, t)\,\tilde{a}_m^*(\bfk', t')\right>\nonumber \\[3ex] 
&=& \widetilde{G}_{\nu}(\bfk,t,0)\,\widetilde{G}_{\nu}(\bfk',t',0)\;\int_0^t \mathrm{d}s\,\int_0^{t'}\mathrm{d}s'\;\times\nonumber \\[2ex]
&&\times\Biggl\{ \left< g_j(\bfk,s)\,g_m^*(\bfk',s') \right> \;+\;\nonumber \\[2ex]
&+& \left[ \Lambda_j(\bfK(\bfk,t))\,-\,\Lambda_j(\bfK(\bfk,s)) \right] \frac{K^2(\bfk,s)}{K_{\bot}^2} 
\left< g_1(\bfk,s)\,g_m^*(\bfk',s') \right> +\nonumber \\[2ex]
&+& \left[ \Lambda_m(\bfK(\bfk',t'))\,-\,\Lambda_m(\bfK(\bfk',s')) \right] \frac{K^2(\bfk',s')}{K_{\bot}^{\prime\,2}} 
\left< g_j(\bfk,s)\,g_1^*(\bfk',s') \right> +\nonumber \\[2ex]
&+& \left[ \Lambda_j(\bfK(\bfk,t))\,-\,\Lambda_j(\bfK(\bfk,s)) \right]\,\left[ \Lambda_m(\bfK(\bfk',t'))\,-\,\Lambda_m(\bfK(\bfk',s')) \right] \times\nonumber\\[2ex] 
&&\qquad\qquad\times\;\frac{K^2(\bfk,s)K^2(\bfk',s')}{K_{\bot}^2K_{\bot}^{\prime\,2}}\left< g_1(\bfk,s)\,g_1^*(\bfk',s') \right>\Biggr\}
\label{fourvvcorr}
\end{eqnarray}

\noindent 
Using equations~(\ref{f-g}) and (\ref{giff}), we write

\begin{eqnarray}
\left< g_j(\bfk,s)\,g_m^*(\bfk',s') \right> &=& \frac{1}{\widetilde{G}_{\nu}(\bfk,s,0)\,\widetilde{G}_{\nu}(\bfk',s',0)}
\left< \tilde{f}_j(\bfK(\bfk,s),s)\,\tilde{f}_m^*(\bfK(\bfk',s'),s') \right>\nonumber\\[2ex]
&=&\frac{1}{\widetilde{G}_{\nu}(\bfk,s,0)\,\widetilde{G}_{\nu}(\bfk',s',0)}(2\pi)^6\,\delta(\bfk-\bfk')\Phi_{jm}(\bfk,s,s') 
\label{ggcorr}
\end{eqnarray}

\noindent 
Using $\widetilde{G}_{\nu}(\bfk,t,0)(\widetilde{G}_{\nu}(\bfk,s,0))^{-1} = \widetilde{G}_{\nu}(\bfk,t,s)$, equations~(\ref{fourvvcorr}), 
(\ref{ggcorr}) and  (\ref{spec}) give, 

\begin{eqnarray}
\Pi_{jm} (\bfk,t,t') &=& \int_0^t \mathrm{d}s\,\int_0^{t'}\mathrm{d}s'\;
\widetilde{G}_{\nu}(\bfk,t,s)\,\widetilde{G}_{\nu}(\bfk,t',s') \times \nonumber \\[2ex]
&&\times\Biggl\{ \Phi_{jm}(\bfk,s,s') \;+\; \nonumber \\[2ex]
&+& \left[ \Lambda_j(\bfK(\bfk,t))\,-\,\Lambda_j(\bfK(\bfk,s)) \right] \frac{K^2(\bfk,s)}{K_{\bot}^2} \Phi_{1m}(\bfk,s,s') \;+\;\nonumber \\[2ex]
&+& \left[ \Lambda_m(\bfK(\bfk,t'))\,-\,\Lambda_m(\bfK(\bfk,s')) \right] \frac{K^2(\bfk,s')}{K_{\bot}^2} \Phi_{j1}(\bfk,s,s') \;+\;\nonumber \\[2ex]
&+& \left[ \Lambda_j(\bfK(\bfk,t))\,-\,\Lambda_j(\bfK(\bfk,s)) \right]\,\left[ \Lambda_m(\bfK(\bfk,t'))\,-\,\Lambda_m(\bfK(\bfk,s')) \right]\times \nonumber \\[2ex]
&& \qquad\qquad\times\;\frac{K^2(\bfk,s)K^2(\bfk,s')}{K_{\bot}^4}\,\Phi_{11}(\bfk,s,s') \Biggr\}
\label{Pi-Phi}
\end{eqnarray}

When $\Phi_{jm} (\bfk,t,t')$ is  real,  the forcing may be called non helical. Then equaton~(\ref{Pi-Phi}) proves that 
the velocity spectrum tensor, $\Pi_{jm} (\bfk,t,t')$ is also a real quantity. In other words, non helical forcing of an 
incompressible fluid at low $\re$, in the absence of Lorentz forces, gives rise to a non helical velocity field. 
In this case, as we noted earlier, the velocity correlators  $Q_{jml}(\bfr,t,t')$ and $C_{jml}(\bfr,t,t')$ are odd functions of $\bfr$ and, 
$G_{\eta}(\bfr,t,t')$ being an even function of $\bfr$, equation~(\ref{trcoef}) implies that the transport coefficient, $\alpha_{il}(\tau)$ 
vanishes. In other words, the $\alpha$--effect is absent for non helical forcing at low $\re$ and $\rem$, for arbitrary values of the shear 
parameter. This may not seem like a particularly surprising conclusion, but it is by no means an obvious one, because at high $\re$ 
it may happen that  $\Pi_{jm} (\bfk,t,t')$ is complex even when $\Phi_{jm}(\bfk,t,t')$ is  real.

We now specialize to the case when the forcing is not only \emph{non helical}, but \emph{isotropic} and \emph{delta--correlated--in--time} 
as well; in this case, 

\beq
\Phi_{jm}(\bfk,s,s') \;=\; \delta(s-s')\,P_{jm}(\bfK(\bfk,s))\,F\left(\frac{K(\bfk,s)}{K_F}\right) 
\label{phidef}
\eeq

\noindent 
where $K(\bfk,s) = \left|\bfK(\bfk,s)\right|\,$, $K_F = \ell^{-1}$ is the wavenumber at which the fluid is stirred, 
$P_{jm}(\bfK) = \left(\delta_{jm} - K_jK_m/K^2\right)$ is a projection operator, and $F(K/K_F)\geq 0$ is the 
\emph{forcing power spectrum}. 

\noindent
Substitute equation~(\ref{phidef}) in (\ref{Pi-Phi}), and reduce the double--time integrals
 to a single--time integral using,

\beq
\int_0^t \mathrm{d}s\,\int_0^{t'}\mathrm{d}s'\,\delta(s-s')\, w(\bfk,s,s')\;=\;\int_0^{t_<}\mathrm{d}s\, w(\bfk,s, s)
\eeq

\noindent
where $t_<\,=\,\text {Min}\,(t,t')$. Then the velocity spectrum tensor, 

\begin{eqnarray}
\Pi_{jm} (\bfk,t,t') &=& \int_0^{t_<}\,\mathrm{d}s\;
\widetilde{G}_{\nu}(\bfk,t,s)\,\widetilde{G}_{\nu}(\bfk,t',s)\,F\left(\frac{K(\bfk, s)}{K_F}\right)\,\times\nonumber \\[2ex]
&&\times\Biggl\{ P_{jm}(\bfK(\bfk,s)) \;+\; \nonumber \\[2ex]
&+&\left[ \Lambda_j(\bfK(\bfk,t))\,-\,\Lambda_j(\bfK(\bfk,s)) \right]\frac{K^2(\bfk,s)}{K_{\bot}^2}P_{1m}(\bfK(\bfk,s)) \;+\; \nonumber \\[2ex]
&+&\left[ \Lambda_m(\bfK(\bfk,t'))\,-\,\Lambda_m(\bfK(\bfk,s)) \right]\frac{K^2(\bfk,s)}{K_{\bot}^2}P_{j1}(\bfK(\bfk,s)) \;+\; \nonumber \\[2ex]
&+&\left[ \Lambda_j(\bfK(\bfk,t))\,-\,\Lambda_j(\bfK(\bfk,s)) \right] \left[ \Lambda_m(\bfK(\bfk,t'))\,-\,\Lambda_m(\bfK(\bfk,s)) \right] \times \nonumber \\[2ex]
&&\qquad\qquad\times\;\frac{K^4(\bfk,s)}{K_{\bot}^4}\,P_{11}(\bfK(\bfk,s))\Biggr\}
\label{Pifinal}
\end{eqnarray}

\noindent
is completely determined when the forcing power spectrum, $F(K/K_F)$, has been specified. 

Let an observer located at the origin of the laboratory frame correlate fluid velocities at time $\tau=t$ and at time $\tau'=t'$. The two--point function that measures this quantity is given by, 

\beq
\left<v_j({\bf 0}, \tau)v_m({\bf 0}, \tau')\right> \;=\; R_{jm}({\bf 0},t,t') \;=\; \int \mathrm{d}^3k\;\Pi_{jm}(\bfk, t, t')
\label{zerocorr}
\eeq

\noindent
It can be proved that, in the long time limit when $t\to\infty$ and $t'\to\infty$, $R_{jm}({\bf 0},t,t')$ is a function only of the time difference, 
$(t-t')$. The \emph{equal--time} correlator, defined by $R_{jm}({\bf 0},t,t)\,$, is symmetric: $R_{jm}({\bf 0},t,t) = R_{mj}({\bf 0},t,t)$. A related quantity is the root--mean--squared velocity, $v_{\rm rms}(t)$, defined by

\beq 
v_{\rm rms}^2(t) \;=\; R_{11}({\bf 0}, t, t) \;+\; R_{22}({\bf 0}, t, t) \;+\; R_{33}({\bf 0}, t, t)
\label{vrms}
\eeq

\noindent
In the long--time limit, both $R_{jm}({\bf 0},t,t)\,$ and $v_{\rm rms}(t)$ saturate due to the balance reached between forcing and viscous dissipation; let $v_{\rm rms}^\infty = \lim_{t\to\infty}v_{\rm rms}(t)$. 

We now define various dimensionless quantities: The \emph{fluid Reynolds number}, $\re = v_{\rm rms}^\infty/(\nu K_F)\,$; 
the \emph{magnetic Reynolds number}, $\rem = v_{\rm rms}^\infty/(\eta K_F)\,$; the \emph{Prandtl number}, 
${\rm Pr} = \nu/\eta\,$; the dimensionless \emph{Shear parameter}, ${\rm S_h} = S/(v_{\rm rms}^\infty K_F)\,$.

For numerical computations, it is necessary to choose a form for the forcing power spectrum. A quite common choice, used especially 
in numerical simulations, is forcing which is confined to a spherical shell of magnitude $K_F$. Therefore, whenever we need to choose a 
form for the forcing power spectrum, we take it to be, 

\beq
F\left(\frac{K}{K_F}\right) \;=\; F_0 \, \delta\left(\frac{K}{K_F} - 1\right)\\[2em] 
\label{Fdef}
\eeq

\section{Predictions and comparison with numerical experiments}

We have already established that the transport coefficient $\alpha_{il} = 0$ when the stirring is non helical. The 
other transport coefficient $\eta_{iml}$ can be calculated by the following steps:

\begin{itemize}

\item[(i)]  
Computing the velocity spectrum tensor, $\Pi_{jm}$, using equations~(\ref{Pifinal}) and (\ref{Fdef}). 

\item[(ii)]
Using this in equation~(\ref{rqdc}) to compute  the velocity correlators $C_{jml}$ and $D_{jm}$. 

\item[(iii)] 
Substituting these correlators in the second of equations~(\ref{trcoef}).

\end{itemize}

\noindent
We also seek to compare our analytical results with measurements of numerical simulations, which use the  test--field 
method \cite{BRRK08}. In this method, the mean--magnetic field is averaged over the coordinates $X_1$ and $X_2$.
So we consider the case when the mean magnetic field, $\bfB =\bfB(X_3, \tau)$.  The condition 
$\bnabla \cendot\bfB = 0$ implies that $B_3$ is uniform in space, and it can be set to zero; hence we have 
$\bfB = (B_1, \,B_2, \,0)$. Thus, equation~(\ref{emfslow}) for the mean EMF gives $\bfemf = ({\cal E}_1, \,{\cal E}_2,\, 0)$, 
with 

\beq
{\cal E}_i \;=\; -\,\eta_{ij}(\tau)\,J_j\,;\qquad\qquad \bfJ \;=\; \bnabla\cross \bfB \;=\; \left( -\frac{\partial B_2}{\partial X_3},
\;\;\frac{\partial B_1}{\partial X_3},\;\;0 \right) 
\label{emftf}
\eeq

\noindent
where $2$--indexed magnetic diffusivity tensor $\eta_{ij}$ has four components, $(\eta_{11}, \,\eta_{12}, \,\eta_{21}, \,\eta_{22})$, 
which are defined in terms of the $3$--indexed object $\eta_{iml}$ by

\beq
\eta_{ij}(\tau) \;=\; \epsilon_{lj3}\,\eta_{i3l}(\tau)\,;\quad\mbox{which implies that}\quad
\eta_{i1}(\tau) \;=\; -\,\eta_{i32}(\tau)\,,\quad\eta_{i2}(\tau) \;=\; \eta_{i31}(\tau)
\label{etatf}
\eeq

\noindent
Equation~(\ref{emftf}) for $\bfemf$ can now be substituted in equation~(\ref{meanindeqn}). Then the 
mean--field induction becomes, 

\begin{eqnarray}
\frac{\partial B_1}{\partial\tau}\,&=&\,-\,\eta_{21}\;\frac{\partial^2 B_2}{\partial X_3^2} \;+\;
\left( \eta + \eta_{22} \right)\frac{\partial^2 B_1}{\partial X_3^2} \nonumber\\[2ex]
\frac{\partial B_2}{\partial\tau}\,&=&\,SB_1\,-\,\eta_{12}\;\frac{\partial^2 B_1}{\partial X_3^2} \;+\;
\left( \eta + \eta_{11} \right)\frac{\partial^2 B_2}{\partial X_3^2}
\label{meanindtf}
\end{eqnarray}

\noindent
The diagonal components, $\eta_{11}(\tau)$ and $\eta_{22}(\tau)$,  augment the microscopic resistivity, $\eta$, 
whereas the off--diagonal components, $\eta_{12}(\tau)$ and $\eta_{21}(\tau)$, lead to cross--coupling of  $B_1$ and $B_2$. 

\subsection{The magnetic diffusivity tensor}

We now use our dynamical theory to calculate  $\eta_{ij}(\tau)$. From equations~(\ref{etatf}) and (\ref{trcoef}), 
we have

\begin{eqnarray}
\eta_{ij}(\tau) &\;=\;& \epsilon_{lj3}\,\eta_{i3l}(\tau)\nonumber\\[3ex]
&\;=\;& \epsilon_{lj3} \epsilon_{ipm}\,\int_0^{\tau} \mathrm{d}\tau' \int \mathrm{d}^3r \;r_3\,
G_{\eta}(\bfr,\tau,\tau')\,C_{pml}(\bfr,\tau,\tau') \;+\;\nonumber\\[2ex]
&&+\; \delta_{ij}\,\int_0^{\tau} \mathrm{d}\tau' \int \mathrm{d}^3r \, G_{\eta}(\bfr,\tau,\tau')\;
D_{33}(\bfr,\tau, \tau') 
\label{etaijcd}
\end{eqnarray}

\noindent
Thus the ``D'' terms contribute only to the diagonal components, $\eta_{11}$ and $\eta_{22}$. This is the expected behaviour of turbulent 
diffusion, which we now see is true for arbitrary shear.  Using equation~(\ref{rqdc}), the velocity correlators $C_{pml}$ and $D_{33}$ can 
now be written in terms of $\Pi_{jm}$. After some lengthy calculations, the $\eta_{ij}(\tau)$ can be expressed in terms of the velocity 
spectrum tensor by, 

\begin{eqnarray}
\eta_{ij}(\tau) &=& 2\eta\int_0^\tau\,\mathrm{d}\tau' \int\,\mathrm{d}^3k\;\widetilde{G}_{\eta}(\bfk,\tau,\tau')\,(\tau-\tau')\, k_3\,
\biggl[\delta_{j2} (k_1 - S\tau' k_2) \,-\, \delta_{j1}k_2\biggr] \times \nonumber \\ [2ex]
&& \qquad\qquad\;\;\times \biggl[\delta_{i1}\left\{ \Pi_{23} - \Pi_{32} - S(\tau-\tau') \Pi_{31} \right\} \,+\,  \delta_{i2} \left\{ \Pi_{31} - 
\Pi_{13}\right \}\biggr] 
\quad+ \nonumber \\[3ex]
&& +\quad \delta_{ij}\,\int_0^\tau\,\mathrm{d}\tau' \int\,\mathrm{d}^3k\;\widetilde{G}_{\eta}(\bfk,\tau,\tau')\,\Pi_{33}
\label{etaij-pi}
\end{eqnarray}

\noindent 
where $\Pi_{lm}\,=\,\Pi_{lm}(\bfk,\tau,\tau')$,  and the indices $(i, j)$ run over values $1$ and $2$. Here $\widetilde{G}_{\eta}(\bfk,\tau,\tau')$
is the Fourier--space resistive Green's function defined in equation~(\ref{frgreta}). The final step in computing $\eta_{ij}(\tau)$ is to use 
equations~(\ref{Pifinal}) and (\ref{Fdef}) for the velocity spectrum tensor, $\Pi_{lm}$. 

\begin{figure}
\includegraphics[scale=0.7,angle=-90]{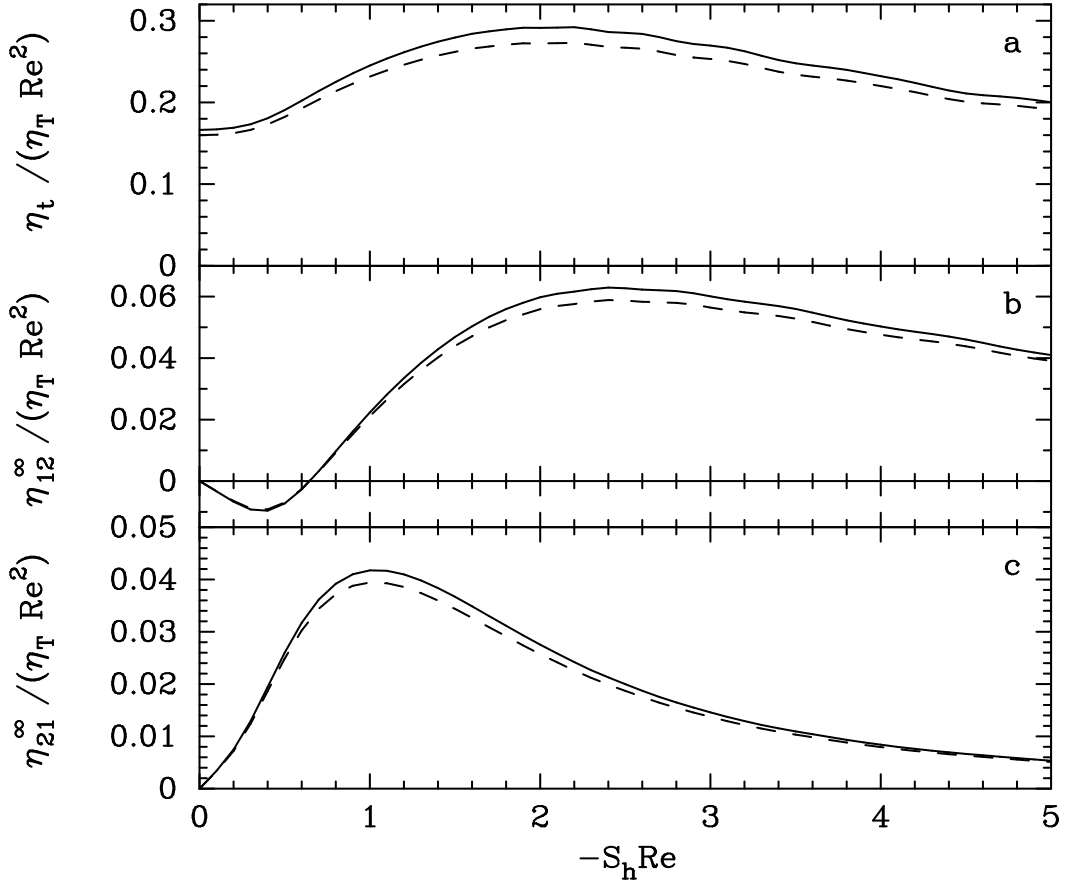}
\caption{Plots of the saturated quantities $\eta_t$, $\eta^{\infty}_{12}$ and $\eta^{\infty}_{21}$ for $\re =\rem =0.1$ and $\re=\rem=0.5$, corresponding to ${\rm Pr} = 1\,$, versus the dimensionless  parameter $\left(-{\rm S_h}\re\right)$. The bold lines are for $\re = \rem = 0.1$, and the dashed lines are for $\re = \rem = 0.5\,$.}
\end{figure}

\begin{figure}
\includegraphics[scale=0.7,angle=-90]{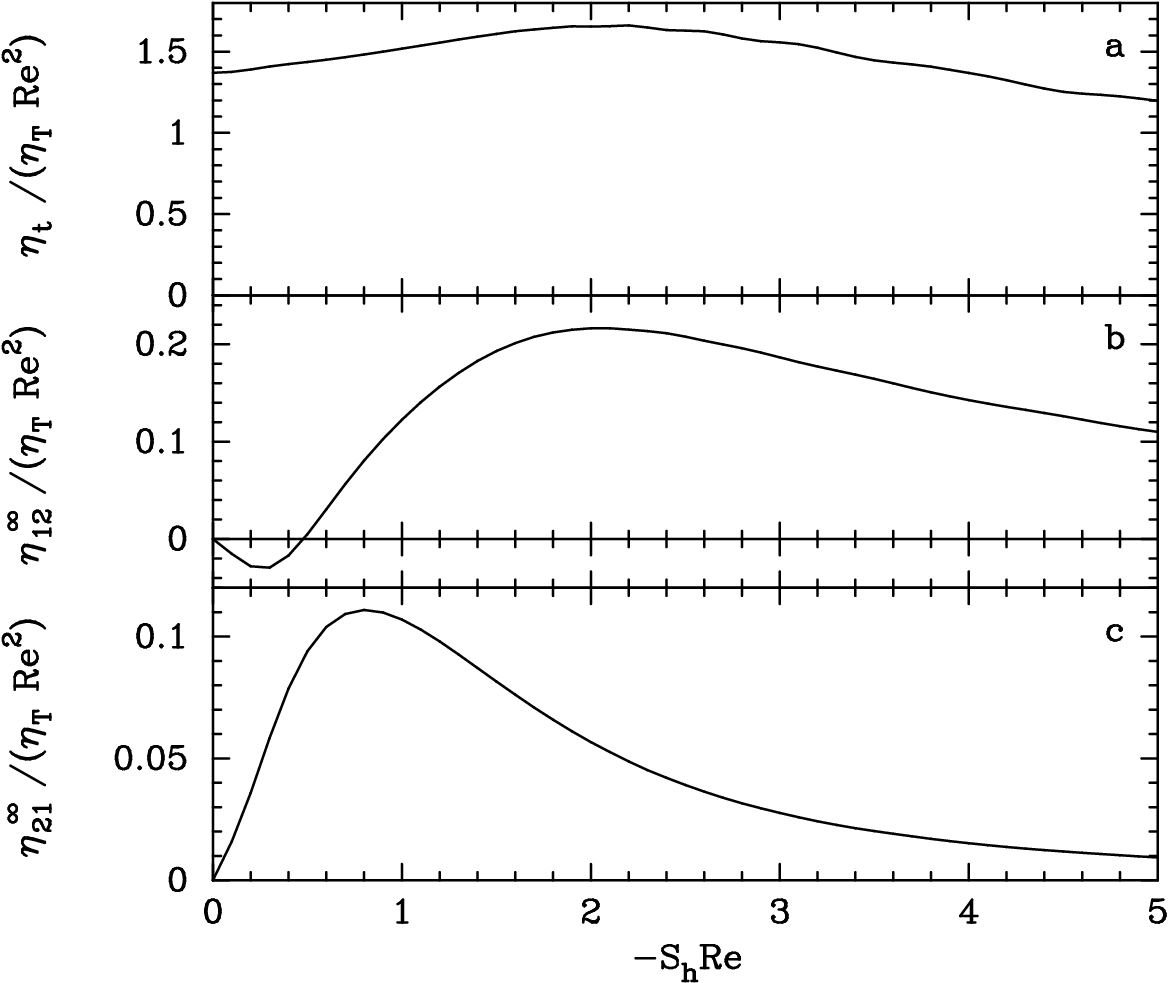}
\caption{Plots of the saturated quantities $\eta_t$, $\eta^{\infty}_{12}$ and $\eta^{\infty}_{21}$ for $\re = 0.1$ and $\rem =0.5$, corresponding to ${\rm Pr} = 5$, versus the dimensionless  parameter $\left(-{\rm S_h}\re\right)$.}
\end{figure}

\begin{figure}
\includegraphics[scale=0.7,angle=-90]{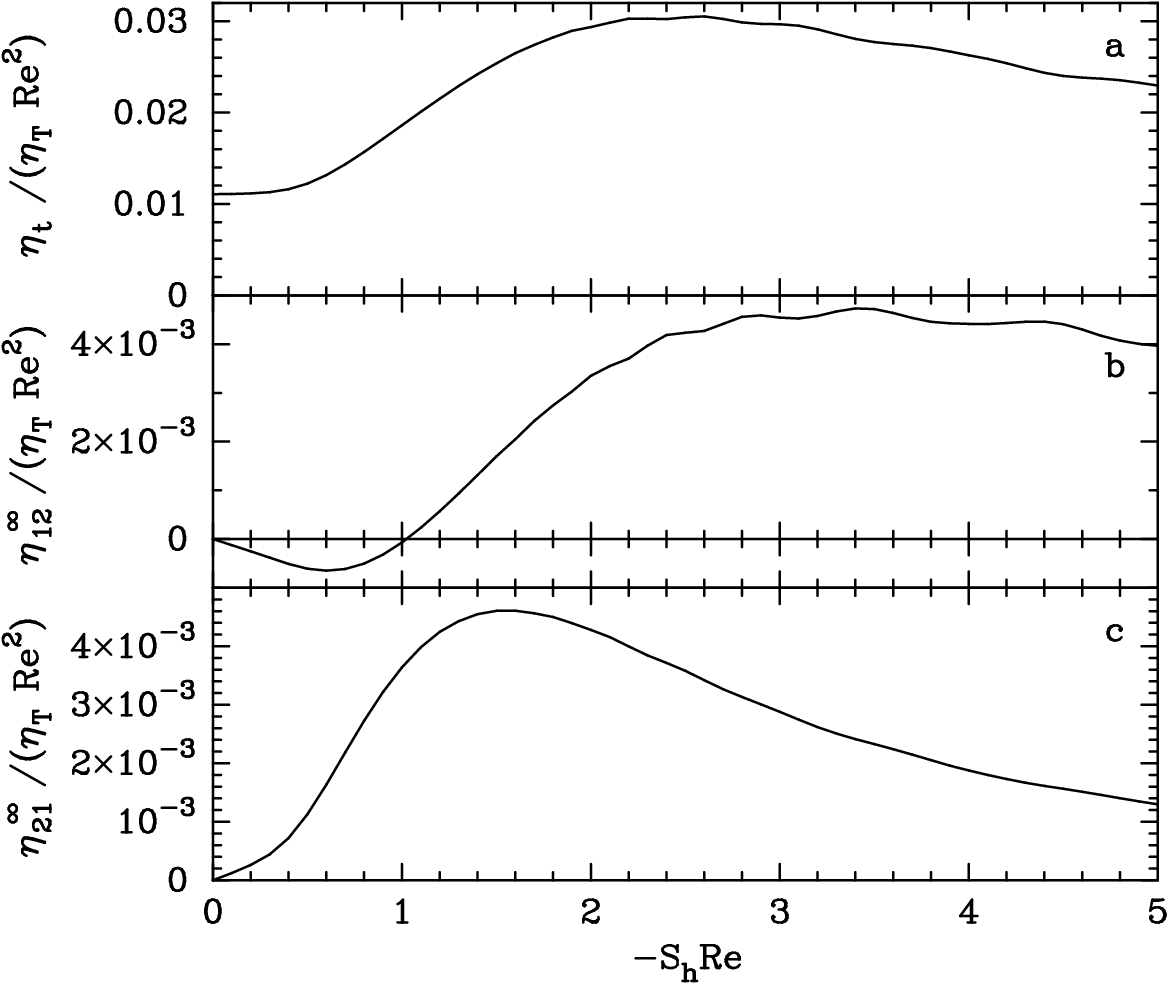}
\caption{Plots of the saturated quantities $\eta_t$, $\eta^{\infty}_{12}$ and $\eta^{\infty}_{21}$ for $\re = 0.5$ and $\rem =0.1$, corresponding to ${\rm Pr} = 0.2$, versus the dimensionless  parameter $\left(-{\rm S_h}\re\right)$.}
\end{figure}

\begin{figure}
\includegraphics[scale=0.7,angle=-90]{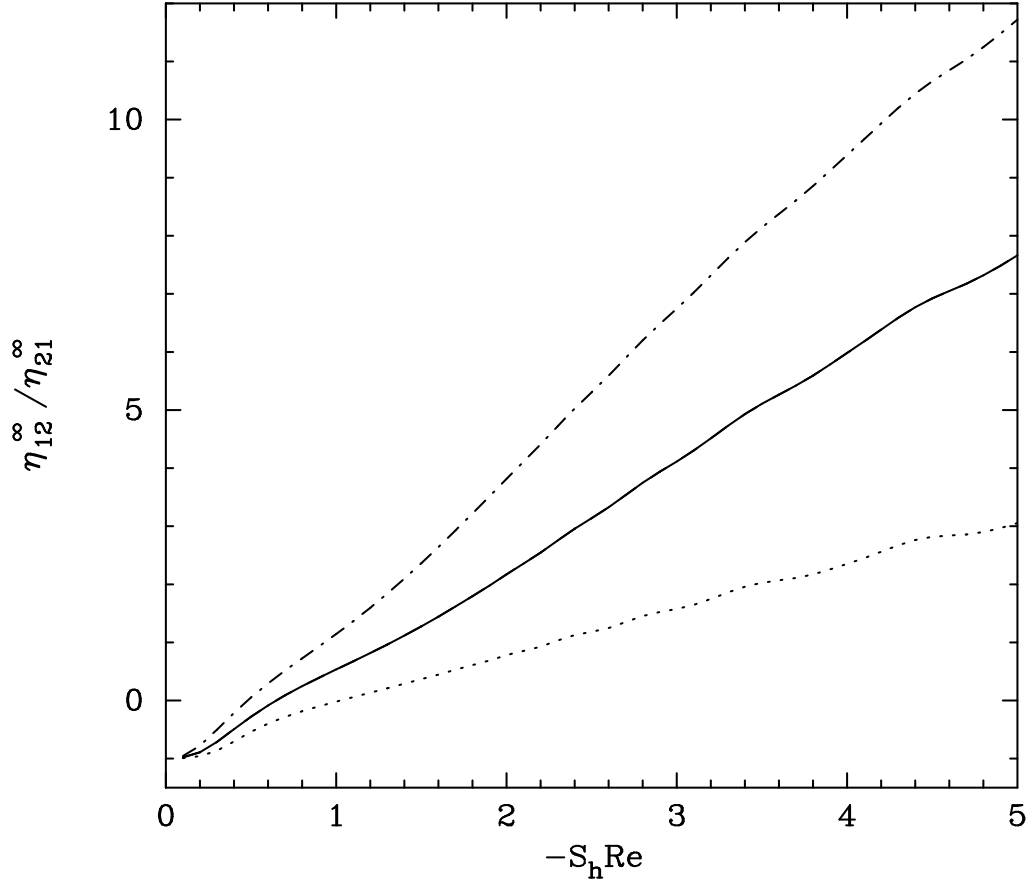}
\caption{Plots of the ratio $\left(\eta^{\infty}_{12}/\eta^{\infty}_{21}\right)$ versus the dimensionless parameter $\left(-{\rm S_h}\re\right)$ for all the cases considered in Figures~(1--3). The bold line is for the two cases corresponding to ${\rm Pr} = 1$, the dashed--dotted line is for  ${\rm Pr} = 5$, and the dotted line is for  ${\rm Pr} = 0.2$.}
\end{figure}

The $\eta_{ij}(\tau)$ saturate at some constant values at late times; let us denote these constant values by $\eta^{\infty}_{ij} = \eta_{ij}(\tau\to\infty)$. If the mean magnetic field  changes over times that are longer than the saturation time, we may use $\eta^{\infty}_{ij}$ instead of the time--varying quantitites $\eta_{ij}(\tau)$ in equation~(\ref{meanindtf}). Looking for solutions $\bfB\propto\exp{\left[\lambda\tau + \mathrm{i} K_3X_3\right]}$, we obtain the dispersion relation,

\beq
\frac{\lambda_{\pm}}{\eta_T \,K_3^2} \;=\; -1\; \pm \, \frac{1}{\eta_T} \sqrt{\eta^{\infty}_{21} 
\left( \frac{S}{K_3^2} + \eta^{\infty}_{12} \right) \,+\,\epsilon^2}
\label{roots}
\eeq

\noindent
given in \cite{BRRK08}, where the new constants are defined as,  

\beq
\eta_t \;=\; \frac{1}{2}(\eta^{\infty}_{11} \,+\, \eta^{\infty}_{22})\;,\qquad \eta_T \;=\; \eta \,+\, \eta_t\;,\qquad 
\epsilon \;=\; \frac{1}{2}(\eta^{\infty}_{11} \,-\, \eta^{\infty}_{22})
\label{etas}
\eeq

\noindent
Exponentially growing solutions for the mean magnetic field are obtained when the radicand in equation~(\ref{roots}) is both positive and exceeds $\eta_T^2\,$. 

From equations~(\ref{etaij-pi}), (\ref{frgreta}), (\ref{Pifinal}) and (\ref{Fdef}), it can be verified that the saturated values of the magnetic diffusivities, $\eta^{\infty}_{ij}$, have the following general functional form:

\beq
\eta^\infty_{ij} \;=\; \eta_T\re^2\frac{f_{ij}({\rm S_h}\re\,, {\rm Pr})}{1 \,+\, \chi({\rm S_h}, \re, {\rm Pr})}\,,
\label{etascalings}
\eeq

\noindent
where the $f_{ij}$ are dimensionless functions of two variables, and $\chi$ is a dimensionless function of three variables. Figures~(1--3) display plots of $\eta_t$,  $\eta^{\infty}_{12}$ and $\eta^{\infty}_{21}\,$, versus the dimensionless parameter $\left(-{\rm S_h}\re\right)$. The scalings of the ordinates have been chosen for compatibility with the functional form displayed in equation~(\ref{etascalings}) above. These plots should be compared with Figure~(3) of \cite{BRRK08}. However, it should be noted that we operate in quite different parameter regimes; we are able to explore larger values of $|{\rm S_h}|$, whereas \cite{BRRK08} have done simulations for larger $\re$ and $\rem$. The plots in Figure~(1a--c) are for ${\rm Pr} = 1$, but for two sets of values of the Reynolds numbers; $\re=\rem = 0.1$, and $\re=\rem = 0.5$. Figure~(2a--c) are for $\re = 0.1$ and $\rem =0.5$, corresponding to ${\rm Pr} = 5$. Figure~(3a--c) are for $\re = 0.5$ and $\rem =0.1$, corresponding to ${\rm Pr} = 0.2$. As may be seen from equation~(\ref{etascalings}), the ratio, $\left(\eta^{\infty}_{12}/\eta^{\infty}_{21}\right)$, is a function only of the two dimensionless parameters, $\left({\rm S_h}\re\right)$ and ${\rm Pr}$. In Figure~(4) we plot this ratio versus $\left(-{\rm S_h}\re\right)$ for all the cases considered in Figures~(1--3). Some noteworthy properties are as follows:

\begin{enumerate}
\item[(i)] We see that $\eta_t$ is always positive. For a fixed value of $\left(-{\rm S_h}\re\right)$, the quantity $\eta_t/(\eta_T\re^2)$ increases with ${\rm Pr}$ and, for a fixed value of ${\rm Pr}$, it increases as $\left(-{\rm S_h}\re\right)$ increases from zero (which is consistent with \cite{BRRK08}), attains a maximum value near $\left(-{\rm S_h}\re\right)\approx 2$, and then decreases while always remaining positive. 

\item[(ii)] 
As expected, the behaviour of $\eta_{12}^{\infty}$ is more complicated. It is zero for $\left(-{\rm S_h}\re\right)=0$, and becomes negative for not too large values of $\left(-{\rm S_h}\re\right)$.  After reaching a minimum value, it then becomes an increasing function of $\left(-{\rm S_h}\re\right)$ and attains positive values for large $\left(-{\rm S_h}\re\right)$. Thus the sign of $\eta_{12}^{\infty}$ is sensitive to the values of the control parameters. This may help reconcile, to some extent, the fact that different signs for $\eta_{12}^{\infty}$ are reported in \cite{RK06} and \cite{BRRK08}. 

\item[(iii)] 
As may be seen, $\eta_{21}^{\infty}$ is always positive. This agrees with the  result obtained in \cite{BRRK08},  \cite{RS06} and \cite{RK06}.  

\item[(iv)] At first sight $\eta_{12}^{\infty}$ and $\eta_{21}^{\infty}$ appear to have quite different behaviour. However, closer inspection reveals 
certain systematics: as ${\rm Pr}$ increases, the overall range of values increases, while their shapes shift leftward to smaller values of $\left(-{\rm S_h}\re\right)$. From equation~(\ref{etascalings}), it 
is clear that the ratio $\left(\eta^{\infty}_{12}/\eta^{\infty}_{21}\right)$ is a function only of the two variables, $\left({\rm S_h}\re\right)$ and ${\rm Pr}$. As Figure~(4) shows, this ratio is nearly a linear function of $\left({\rm S_h}\re\right)$, whose slope increases with ${\rm Pr}$.

\item[(v)]
The magnitude of the quantity, $\chi({\rm S_h}, \re, {\rm Pr})$, that appears  in equation~(\ref{etascalings}), is much smaller than unity. So $\eta_t/(\eta_T\re^2)$, $\eta_{12}^{\infty}/(\eta_T\re^2)$ and $\eta_{21}^{\infty}/(\eta_T\re^2)$ can be thought of (approximately) as functions of $\left(-{\rm S_h}\re\right)$ and ${\rm Pr}$. This is the reason why, in Figure~(1), the bold and dashed lines lie very nearly on top of each other. 

\end{enumerate}

\subsection{Implications for dynamo action and the shear--current effect}

The mean magnetic field has a growing mode if the roots of equation~(\ref{roots}) have a positive real part. It is clear that the real part of $\lambda_-$ is always negative. So, for the growth of the mean magnetic field, the real part of $\lambda_+$ must be positive. Requiring this, we see from equation~(\ref{roots}) that the condition for dynamo action is, 

\beq
\frac{\eta^{\infty}_{21}S}{\eta_T^2K_3^2} \;+\; \frac{\eta^{\infty}_{12}\eta^{\infty}_{21}}{\eta_T^2} \;+\; \frac{\epsilon^2}{\eta_T^2} \;>\; 1.
\label{dyncond}
\eeq 

\noindent
In Figure~(5) we plot the last two terms, $\left(\eta^{\infty}_{12}\eta^{\infty}_{21}/\eta_T^2\right)$ and $\left(\epsilon^2/\eta_T^2\right)$, as functions of $\left(-{\rm S_h}\re\right)$, for all the four cases, $\re = \rem = 0.1\,$; $\re = \rem = 0.5\,$; $\re = 0.1, \rem = 0.5\,$ and $\re = 0.5, \rem = 0.1\,$. As may be seen, the magnitudes of both terms are much smaller than unity, so they are almost irrelevant for dynamo action. Hence, 
there is growth of the mean magnetic field only when the first term, $\left(\eta^{\infty}_{21}S/\eta_T^2K_3^2\right)\,$, exceeds unity. This is possible for small enough $K_3^2$, so long as $\left(\eta^{\infty}_{21}S\right)$ is positive. However, we see from Figures~(1--3) that $\eta^{\infty}_{21}$ is always positive, implying that the product $\left(\eta^{\infty}_{21}S\right)$ is always negative. Therefore the inequality of (\ref{dyncond}) cannot be satisfied, and the mean--magnetic field always decays, a conclusion which is in agreement with those of  \cite{BRRK08}, \cite{RS06} and \cite{RK06}. We can understand the above results more physically. Let us assume that $\left|K_3\right|$ is small enough, and keep only the most important terms
in equation~(\ref{meanindtf}). Then we have, 

\beq
\frac{\partial B_1}{\partial\tau} \;=\;-\,\eta_{21}^\infty\;\frac{\partial^2 B_2}{\partial X_3^2} \;+\; \dots\,,\qquad\quad \frac{\partial B_2}{\partial\tau} \;=\; SB_1 \;+\; \dots\,,
\label{shcurexp}
\eeq

\noindent
where we have used the saturated values of the magnetic diffusivity. If we now look for modes of the form $\bfB\propto\exp{\left[\lambda\tau + \mathrm{i} K_3X_3\right]}$, we obtain the dispersion relation, $\lambda_\pm = \pm\,K_3\sqrt{\eta^{\infty}_{21}S}$. So it is immediately obvious that $\lambda_+$ is real and positive --- i.e. the mean magnetic field grows --- only when the product $\left(\eta^{\infty}_{21}S\right)$ is positive. However, this product happens to be negative, and the mean magnetic field is a decaying wave.

\begin{figure}
\includegraphics[scale=0.7,angle=-90]{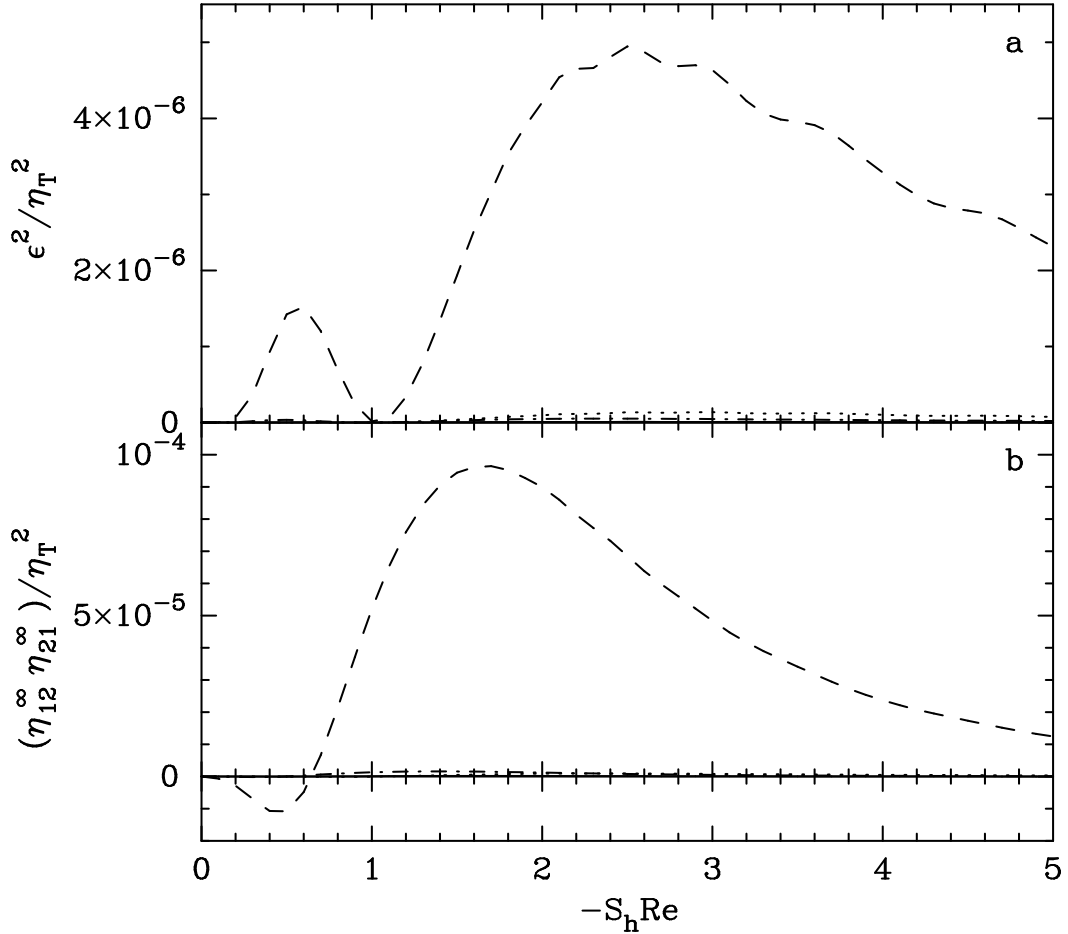}
\caption{Plots of $\left(\epsilon^2/\eta_T^2\right)$ and $\left(\eta^{\infty}_{12}\eta^{\infty}_{21}/\eta_T^2\right)$ versus the dimensionless parameter $\left(-{\rm S_h}\re\right)$ for all the four cases considered in Figures~(1--3). The bold lines are for $\re = \rem = 0.1\,$; the dashed lines are for $\re = \rem = 0.5\,$; the dashed--dotted lines are for $\re = 0.1, \rem = 0.5\,$ and the dotted lines are for $\re = 0.5, \rem = 0.1\,$.} 
\end{figure} 

The above results have direct bearing on the \emph{shear--current effect} \cite{RK03}. This effect refers to an extra contribution to the mean EMF which is perpendicular to both the mean vorticity (of the background shear flow) and the mean current. From equation~(\ref{emftf}), we see that, in our case, the relevant term is the contribution, $-\eta_{21}^\infty J_1$, to ${\cal E}_2$. As Figures~(1--3) show, the diffusivity, $\eta_{21}^\infty$ is non zero only in the presence of shear, so the word \emph{shear} refers to this. The word \emph{current} refers to $J_1$, the cross--field component of the electric current associated with the mean--magnetic field \footnote{Shear also makes an additional contribution through the $SB_1$ contribution to 
$(\partial B_2/\partial\tau)$, which accounts for the product $\left(\eta^{\infty}_{21}S\right)$ playing an important role. However, this is just the well--known physical effect of the shearing of cross-shear component of the mean magnetic field to generate a shear--wise component; it does not have any bearing on the word \emph{shear} in the phrase \emph{shear--current effect}.}. The shear--current effect would lead to the growth of the mean magnetic field (for small enough $K_3$), if only the product $\left(\eta^{\infty}_{21}S\right)$ is positive. However, as we have demonstrated, this product is negative, so the shear--current effect cannot be responsible for dynamo action, at least for small $\re$ and $\rem$, but for all values of the shear parameter.

\section{Conclusions}

Building on the formulation of \cite{SS10}, we have developed a theory of
the \emph{shear dynamo problem} for small magnetic and fluid Reynolds numbers, but for arbitrary values of the shear 
parameter. Our primary goal is to derive precise analytic results which can serve as benchmarks for comparisons with
numerical simulations. A related goal is to resolve the controversy surrounding the nature of the shear--current effect,
without treating the shear as a small parameter.  We began with the expression for the Galilean--invariant mean EMF 
derived in \cite{SS10}, and specialized to the case of a mean magnetic field that is slowly varying in time. This resulted in the 
simplification of the mean--field induction equation, from an integro--differential equation to a partial differential equation. 
This reduction is the first step to the later comparison with the numerical experiments of \cite{BRRK08}.
Explicit expressions for the transport coefficients, $\alpha_{il}$ and $\eta_{iml}$, were derived in terms of the two--point 
velocity correlators which, using results from \cite{SS10}, were then expressed  in terms of the velocity spectrum tensor. 
Then we proved that, when the velocity field is non helical, the transport coefficient  $\alpha_{il}$ vanishes; just like  everything 
else in this paper, this result is non perturbative in the shear parameter. We then considered forced, stochastic dynamics 
for the incompressible velocity field at low Reynolds number. An exact, explicit solution for the velocity field was derived, 
and the velocity spectrum tensor was calculated in terms of the Galilean--invariant forcing statistics. For non helical forcing, the 
velocity field is also non helical and the transport coefficient  $\alpha_{il}$ vanishes, as noted above. We  then specialized to the 
case when the forcing is not only non helical, but isotropic and delta--correlated--in--time as well. We considered the case when the mean--field was a function only of the spatial coordinate $X_3$ and time $\tau\,$; the purpose of this simplification was to facilitate comparison with the  numerical experiments of \cite{BRRK08}. Explicit expressions were derived for
all four components, $\eta_{11}(\tau)$,  $\eta_{22}(\tau)$ $\eta_{12}(\tau)$ and $\eta_{21}(\tau)$, of the magnetic diffusivity tensor, in terms of the  velocity spectrum tensor. Important properties of this fundamental object are as follows:

\begin{enumerate}
 
\item  
All the components of $\eta_{ij}$ are zero at $\tau=0$, and saturate at finite values at late times, which we denote by $\eta^\infty_{ij}\,$.

\item
The off--diagonal components, $\eta_{12}$ and $\eta_{21}$, vanish when the microscopic resistivity vanishes.
  
\item
The sign of $\eta_{12}^{\infty}$ is sensitive to the values of the control parameters. This may help reconcile, to some extent, the fact that different signs for $\eta_{12}^{\infty}$ are reported in \cite{RK06} and \cite{BRRK08}.

\end{enumerate}

We derived the condition --- the inequality~(\ref{dyncond}) --- required 
for the growth of the mean magnetic field: the sum of three terms must exceed unity. It was demonstrated that two of the terms are very small in magnitude, and hence dynamo action was controlled by the behaviour of one term. i.e. 
the mean magnetic field would grow if $\left(\eta^{\infty}_{21}S/\eta_T^2K_3^2\right)$ exceeds unity. This is possible for small enough $K_3^2$, so long as $\left(\eta^{\infty}_{21}S\right)$ is positive. However, we see from Figures~(1--3) that $\eta^{\infty}_{21}$ is always positive, implying that the product $\left(\eta^{\infty}_{21}S\right)$ is always negative. Thus the mean--magnetic field always decays, a conclusion which is in agreement with those of  \cite{BRRK08}, \cite{RS06} and \cite{RK06}.
We then related the above conclusions to the shear--current effect, and demonstrated that the shear--current effect cannot be responsible for dynamo action, at least for small $\re$ and $\rem$, but for all values of the shear parameter. In \cite{BRRK08}, it is suggested that the dynamo action observed in their numerical experiments  might be due to a fluctuating $\alpha$--effect; addressing this issue is being the scope of our present calculations.

\end{document}